\documentstyle [emulateapj]{article}

\newcommand{\etal}{ et al. \,}
\newcommand{\kms}{km~s$^{-1}$ \,}

\def\gtsim{\ {\raise-0.5ex\hbox{$\buildrel>\over\sim$}}\ }
\def\ltsim{\ {\raise-0.5ex\hbox{$\buildrel<\over\sim$}}\ }

\lefthead{Lee and Byun}
\righthead{Dwarf Galaxy UKS 2323--326}

\begin{document}

\title{Stellar Populations of the Dwarf Galaxy UKS 2323--326 in the Sculptor Group
}

\author{Myung Gyoon Lee}
\affil{Department of Astronomy, Seoul National University, Seoul 151-742, 
Korea \\
Electronic mail: mglee@astrog.snu.ac.kr}

\and

\author{Yong-Ik Byun}
\affil{Department of Astronomy and Center for Space Astrophysics, Yonsei University, Seoul 120-749, Korea \\
Electronic mail: byun@darksky.yonsei.ac.kr}




\begin{abstract}

We present deep $BVRI$ CCD photometry of the stars in the dwarf irregular
galaxy UKS 2323--326 in the Sculptor Group. 
The color-magnitude diagrams of the measured stars in UKS 2323--326 
show a blue plume which consists mostly of young stellar
populations, and a well-defined red giant branch (RGB). 
The tip of the RGB is found to be at $I_{\rm TRGB} = 22.65 \pm 0.10$ mag. 
From this the distance to this galaxy
 is estimated to be $d = 2.08\pm 0.12 $ Mpc.
 The corresponding distance of this galaxy
from the center of the Local Group is 1.92 Mpc, showing that it is located outside  the Local Group. The large distance, combined with its velocity
information, indicates that it is very likely to be a member of the Sculptor Group.
The mean metallicity of the red giant branch is estimated to be
[Fe/H] = $-1.98\pm0.17$ dex. 
Total magnitudes of UKS 2323--326 ($<r_H = 70$ arcsec)
 are derived to be $B^T=14.07$ mag, $V^T=13.50$ mag, $R^T=13.18$ mag, and $I^T=12.83$ mag, 
and the corresponding absolute magnitudes
are $M_B=-12.58$ mag, $M_V=-13.14$ mag,  $M_R=-13.45$ mag,
and $M_I=-13.79$ mag.
Surface brightness profiles of the central part of UKS 2323--326 are approximately 
consistent with a King model with a core concentration parameter
$c = \log (r_t / r_c ) \approx 1.0$, and those of the outer part follow
an exponential law with a scale length of 21 arcsec.
The magnitudes and colors of the brightest blue and red stars in UKS 2323--326 (BSG and RSG) are measured to be, respectively, 
$<V(3)>_{BSG} = 20.33\pm 0.25$ mag, $<(B-V)(3)>_{BSG} = 0.14\pm 0.07$ mag, $<V(3)>_{RSG} = 20.74\pm 0.18$ mag, and $<(B-V)(3)>_{RSG} = 1.35\pm 0.08$ mag.
The corresponding absolute magnitudes are derived to be 
$<M_V(3)>_{BSG} = -6.31$ mag and $<M_V(3)>_{RSG} = -5.91$ mag, which
are about half magnitude fainter than those 
expected from conventional correlations with galaxy luminosity.

\centerline{[To appear in the Astronomical Journal in August, 1999]}
\vskip 0.1in  
\end{abstract}


\keywords{galaxy: evolution --- galaxies: irregular --- 
galaxies: individual (UKS 2323--326) --- galaxies: stellar content --- 
galaxies: photometry --- Distance scale}


%

\section{INTRODUCTION}

UKS 2323--326 (UGCA 438) is a faint dwarf irregular 
galaxy in Sculptor discovered 
by Longmore \etal (1978). In the discovery paper, Longmore \etal estimated
the distance to this galaxy from the rough estimate of the 
magnitude of the brightest stars,
to be 1.3 Mpc with a large error of $\pm50$ \%. 
Since then this galaxy has been often considered to be a member 
of the Local Group (\cite{lon78}, \cite{mat98}). 
On the other hand, 
the velocity of UKS 2323--326 from the center of
the Local Group, 82 \kms (\cite{lon78}), suggests that it may not belong to the Local Group (\cite{van94}, \cite{lee95a}). 
Accurate measurement of the distance to this galaxy is needed to resolve
this problem. 

Longmore \etal (1978) measured the HI flux of $15\pm3$ Jy \kms at the 
heliocentric velocity of $62 \pm 5$ \kms,
 and derived, using eye-estimate with the photographic plates, 
the magnitude of the brightest blue stars in UKS 2323--326, 
$B = 19.3\pm0.5$ mag.
 After the discovery paper, however, there has been published 
no detailed photometric study for this galaxy.  

In this paper we present a study of stellar populations of UKS 2323--326 
based on deep $BVRI$ CCD photometry, and 
show that this galaxy is outside the Local Group and is a member of the
Sculptor Group.  
This paper is composed as follows.
Section 2 describes the observations and data reduction, and
Section 3 investigates the morphological structure of UKS 2323--326. 
Section 4 presents the color-magnitude diagrams of UKS 2323--326, 
Section 5 estimates the distance to UKS 2323--326.
Section 6 presents the surface photometry of UKS 2323--326
and Section 7 discusses the group membership, 
 the stellar populations, and the brightest stars of UKS 2323--326.
Finally, summary and conclusion are given in  Section 8.

\section{OBSERVATIONS AND DATA REDUCTION}

$BVRI$ CCD images of UKS 2323--326 were obtained on the photometric night of 1994 October 7 (UT) using the University of Hawaii
2.2 m telescope at Mauna Kea.
Table 1 lists the journal of the observations of UKS 2323--326.
A grey scale map of the $V$-band CCD image of UKS 2323--326 
is displayed in Fig. 1. The size of the field of view is $7'.5 \times 7'.5$
and the ($2\times 2$ binned) pixel scale of the CCD is 0.44 arcsec pixel$^{-1}$.

For the analysis of the data 
we have divided the field covered by our CCD images into three regions 
as shown in Fig. 1:
the C-region which covers the central region ($r<44''$) of UKS 2323--326, 
the I-region which covers the outer region ($44''<r<77''$) of the galaxy , 
and the  F-region which represents
a control field with the same area of the C-region plus I-region. 

Instrumental magnitudes of the stars in the CCD images were derived using
DoPHOT (\cite{sch93}).
These magnitudes were transformed onto the standard system using the
standard stars observed during two nights 
including the same night (\cite{lan92}). 
The transformation equations we derived from the photometry of the standard
stars are: 
$V = v - 0.075 (b-v) -0.118 X + constant$,
$(B-V) = 1.130 (b-v) -0.111 X + constant$,
$(V-R) = 0.970 (v-r) -0.030 X + constant$, and
$I = i +0.050 (v-i) - 0.082 X + constant$,
where upper cases and lower cases represent, respectively, the standard
system and instrumental system. $X$ represents the air mass.
The rms scatter of the solutions are 0.01 -- 0.02 mag.
The total number of stars which were measured at $V$ and at least one other
color in the CCD image is $\sim$2,000. 
Table 2 lists $BVRI$ photometry of the measured
bright stars with $V<22.5$ mag in the field of UKS 2323--326. 
The coordinates X and Y in Table 2 are given in units of 
pixel ( = 0.44 arcsec), and increase to the east and to the south, respectively.

\section{MORPHOLOGICAL STRUCTURE}

Unfortunately there is a very bright star 
$20''$ south-east of the center of UKS 2323--326, as shown in Fig. 1. 
It is about 14 mag foreground star (\cite{lon78}) and is saturated in 
all our CCD images. 
Fig.1 shows that there are several tens of bright stars concentrated in the
central region of UKS 2323--326. 
These stars are considered mostly to be the members
of UKS 2323--326 and are young massive stars as shown later. 
The main body of the galaxy is mostly seen inside the C-region, and  
is slightly elongated along  the south-east direction. 
However, the outer part of the galaxy seen
at the faint level extends out to the boundary of the I-region ($r=77''$), and
is impressively almost circular. There are few bright stars in the I-region.
Instead there are many faint stars seen better in $R$ and $I$ images of the 
I-region, which are probably old red giant stars as seen later. 
Therefore UKS 2323--326 is a circular galaxy, in the central region of which 
there are irregularly distributed some bright stars. 
This kind of structure that some young stars are irregularly distributed
against the smooth background of old stellar populations is common among
irregular galaxies (\cite{san71}, \cite{lee93b}, \cite{lee93}, \cite{min96}).

\section{COLOR-MAGNITUDE DIAGRAMS}

UKS 2323--326 is located 71 deg below the galactic plane in the sky
so that the foreground reddening
for UKS 2323--326 is expected to be very low.
We adopt in this study the foreground reddening value of
$E(B-V)=0.014$  mag for UKS 2323--326 given by Schlegel, Finkbeiner, \& Davis
 (1998).

We display $V$--$(B-V)$ and $I$--$(V-I)$ diagrams of 
 520 measured stars in the C-region,
 280 measured stars in the I-region, and 89 measured stars in the 
F-region in Figs. 2 and 3. 
Note that the area of the (C+I)-region in the field is the same 
as that of the F-region so that we can estimate the contamination due to 
foreground stars by comparing directly the diagrams of each region.

Several distinguishable features of the stars in UKS 2323--326 are seen in 
Figs. 2 and 3.
First, there is a blue plume of bright stars with $(B-V)<0.4$ in the
C-region (filled circles). 
Comparison of the C-region and the F-region in Fig. 2 
shows that these bright blue stars  are mostly  members of UKS 2323--326.
The brightest end of the blue plume extends up to $V\approx 20.0$ mag and
$(B-V)\approx -0.1$. These stars are mostly massive stars which were formed
recently. 
Stars in the I-region are all fainter than $V=23$ mag (open circles), 
much fainter than those bright stars in the C-region. 
This shows that there was little star formation in the outer part
of UKS 2323--326 recently.

Secondly, there are five red bright stars with $20.0<V<21.5$ mag 
and $1.2<(B-V)<1.5$ in the C-region in Fig. 2. This area of the F-region 
color-magnitude diagram is devoid of stars.
Therefore these stars are probably the members of UKS 2323--326.
These stars are considered to be red supergiant stars which were formed recently. 

Thirdly, Fig. 3 shows that there is a strong concentration of red stars fainter
than $I \approx 21.5$ mag in the (C+I)-region. Most of these stars are
probably old red giant branch (RGB) stars of UKS 2323--326.

\section{DISTANCE AND METALLICITY}

We estimate the distance to UKS 2323--326 using the $I$ magnitude of the tip
of the RGB (TRGB), as 
described in Da Costa \& Armandroff (1990) and Lee, Freedman, \& Madore (1993). 
The $I$ magnitude of the TRGB is estimated
using the $I-(V-I)$ diagram in Fig. 3 and 
the luminosity function of red giant stars.
Fig. 4 shows the $I$-band luminosity function of the measured red stars 
in the I-region (solid line) and the C-region (dotted line), from which
the contribution due to field stars was subtracted. 
We also plot the luminosity functions of the similar stars
in the field with the same area as the (C+I) region in Fig. 4, which shows
that the contribution due to field stars is very small.

The I-region is more appropriate for measuring the magnitude of the TRGB
than the C-region, because there are several young bright stars in the C-region
and because the crowding is more severe in the C-region than in the I-region. 
In Fig. 4, as the magnitude increases, there is a sudden
increase at $I=22.65\pm0.10$ mag in the luminosity function of the
I-region, which corresponds to the TRGB seen in the color-magnitude 
diagram in Fig. 3. Several stars brighter than the TRGB are mostly
asymptotic giant stars of intermediate age, also seen in other dwarf
galaxies (\cite{lee95b}). A dip seen at the magnitude of the TRGB in the
luminosity function of the C-region is a statistical fluke. 

The mean color of the TRGB is estimated
to be $(V-I)=1.43\pm 0.05$.
The bolometric magnitude of the TRGB is then calculated from
$M_{\rm bol}=-0.19{\rm [Fe/H]} - 3.81$.
Adopting a mean metallicity of [Fe/H] = $-1.98\pm 0.17$  dex as 
estimated below, we obtain a value for 
the bolometric magnitude of $M_{\rm bol}=-3.43$ mag.
The bolometric correction at $I$ for the TRGB is estimated 
to be BC$_I=0.54$ mag,
adopting a formula for the bolometric correction 
BC$_I$ = 0.881 -- 0.243$(V-I)_{\rm TRGB}$.
The intrinsic $I$ magnitude of the TRGB is then given by
$M_I= M_{\rm bol} - {\rm BC}_I = -3.97$ mag.
Finally the distance modulus of UKS 2323--326 is obtained:
$(m-M)_0 = 26.59 \pm 0.12$ mag 
(corresponding to a distance of $2.08\pm 0.12$ Mpc)
for an adopted extinction of $A_I=0.03$ mag.

We have estimated the mean metallicity of the RGB stars in UKS 2323--326 
using  the $(V-I)$ color of the stars 0.5 mag fainter than the TRGB, $(V-I)_{-3.5}$.
This color is measured from the median value of the colors of 17 red giant
branch stars with $I=23.10\pm0.10$ mag, to be $(V-I)_{-3.5} = 1.29 \pm 0.04$.
From this value
we estimate the mean metallicity to be [Fe/H] $= -1.98\pm 0.17$ dex.
In Fig. 5 we overlayed the loci of the red giant branches of
Galactic globular clusters, of M15, M2, and NGC 1851, shifted according to
the distance and reddening of UKS 2323--326.  
The metallicities of M15, M2 and NGC 1851 are 
[Fe/H] = --2.17, --1.58 and --1.29 dex, respectively.
Fig. 5 shows that the bright part of the RGB of UKS 2323--326
 is located well between those of M15 and M2. 
The broadening of the faint part of the RGB is mostly
due to the photometric errors, as shown by the error bars in Fig. 5.
Thus the mean metallicity of the RGB stars in UKS 2323--326 is very low,
and is close to the lowest end in the metallicity of dwarf irregular galaxies (\cite{lee93}; Lee 1995a,b;\cite{mat98};\cite{lee99}).


\section{SURFACE PHOTOMETRY}

It is very difficult to derive reliably the surface photometry of UKS 2323--326,
because of the presence of
very bright foreground star close to the center of the galaxy.
We have obtained  the surface photometry of UKS 2323--326 as follows.
First, we removed in the original CCD images the images of
very bright foreground star and  several other 
bright stars which were obviously considered to be
foreground stars in the area of UKS 2323--326, using IMEDIT in IRAF. 
Then we performed aperture photometry of 
UKS 2323--326 using the circular annular aperture.
The value of the sky background was estimated from the mean intensity
of the F-region.
Subtraction of the very bright 14 mag star from the image of 
the galaxy is very difficult. The error due to this problem
is not easy to quantify, but may be significant in the photometry
of the galaxy.
Therefore the surface photometry of the galaxy derived in this study
is only approximate. 

The results of surface photometry of UKS 2323--326 are listed in Table 3 and 
are displayed in Fig. 6. In Table 3 $r_{\rm eff}$ presents the mean major radius
of an annular aperture, and $r_{\rm out}$ represents the outer radius of an annular aperture.
Fig. 6(a) shows that the surface brightness profiles are almost
flat in the central region of the galaxy ($r<20$ arcsec), 
and follow approximately the exponential law 
in the outer part ($r>20$ arcsec).
We fit the $V$ surface brightness profile of the inner region of UKS 2323--326 
with the single-mass isotropic King models (\cite{kin66}) and that
of the outer region with the exponential law in Fig. 7.
Fig. 7 shows that the surface brightness profile of the inner region
is roughly fit by a King model with a core concentration parameter
$c = \log ( r_t / r_c ) \approx 1.0$, where $r_c$ and $r_t$ represent the core radius and tidal radius, respectively.
The surface brightness profiles of the outer region
are fit roughly by an exponential law 
with a scale length of $r=21$ arcsec = 210 pc. 
Fig. 6(b) illustrates the surface color profiles for the region at $r<40$ arcsec.
The colors files are plotted only for the region at $r<40$ arcsec, beyond which
the errors of the colors are too large.
The color profiles are almost constant and 
become slowly redder outward from the center to $r\approx 40$ arcsec. 
The rapid change of the color profiles right close
to the center is due to a bright star in the central region.
The properties of the surface brightness and color profiles of UKS2323--326 
are similar to those of other dwarf galaxies
(\cite{kor89}, \cite{mat98}).

From the data of the surface photometry we have derived several
basic parameters of UKS 2323--326 as follows.
The standard radius and Holmberg radius of UKS 2323--326 are measured to be
$r_{25}$ = 46 arcsec = 460 pc and $r_H$ = 70 arcsec = 700 pc, respectively.
These radii are similar, respectively, to the radii of the C-region and 
I-region shown in Fig. 1. 
Note that our value for the Holmberg radius is much larger than
the value Longmore \etal (1978) derived roughly from their photographic plate,
$r_H$ = 45 arcsec.

The central surface brightness of UKS 2323--326 is measured to be 
 $\mu_B (0) = 23.4 $  mag arcsec$^{-2}$, 
$\mu_V (0) =22.9  $ mag arcsec$^{-2}$,
 $\mu_R (0) = 22.6 $  mag arcsec$^{-2}$, and
$\mu_I (0) =22.2  $ mag arcsec$^{-2}$.
The total magnitudes of UKS 2323--326 within $r_H$ are derived to be
$B^T=14.07$ mag, $V^T=13.50$ mag, $R^T=13.18$ mag, and $I^T=12.83$ mag, 
and the corresponding absolute magnitudes
are $M_B=-12.58$ mag, $M_V=-13.14$ mag,  $M_R=-13.45$ mag,
and $M_I=-13.79$ mag.
$B$ total magnitude of UKS 2323--326 derived in this study, $B^T=14.07$ mag, 
is about 1 mag brighter than that given by Longmore \etal (1978), $B^T=15.2$ mag.

\section{DISCUSSION}

\subsection{The Group Membership of UKS 2323--326}

We have measured the distance to UKS 2323--326 to be $d=2.08\pm0.12$ Mpc
 from the
$I$-band magnitude of the TRGB, which is an accurate distance indicator
for resolved galaxies (\cite{lee93}, \cite{sal98}).
Our value is 1.6 times larger and much more accurate than the value
 Longmore \etal (1978) suggested, from the eye-estimate of the magnitude
of the blue brightest stars, $d = 1.3$ Mpc with an error of $\pm 50$ \%.   
Our result shows that UKS 2323--326 is obviously outside the Local Group. 
The mean distance of the Sculptor group is 2.5 Mpc and the radius of the
Sculptor group is $\sim$1.1 Mpc (\cite{puc88}, \cite{cot97}, \cite{jer98}).
UKS2323--326 is located in the western boundary of the Sculptor group
in the sky (see the map of the Sculptor group members given in Fig. 4 of
\cite{cot97}).
Therefore UKS 2323--326 is definitely considered to be a member of the Sculptor
group. 

This conclusion is consistent with the measured velocity of UKS 2323--326. 
From the measured heliocentric velocity of this galaxy, $v=62\pm5$ \kms, the velocity of UKS 2323--326 from the center of
the Local Group is calculated to be $84\pm 5$ \kms ~(\cite{lon78}, \cite{lee95b}). 
This value is 24 \kms ~ larger than the maximum value of 60 \kms 
typically used as the boundary of the Local Group, showing that this galaxy
is probably outside the boundary of the Local Group.

C\^ote \etal (1997) excluded UKS 2323-326 in the list of dwarf candidates
for the Sculptor group, based on the results given by Longmore \etal (1978).
However, our results show that this galaxy should be included as a member
of the Sculptor group.

\subsection{Stellar Populations in UKS 2323--326}

In the discovery paper  Longmore \etal (1978) detected no HII region in 
UKS 2323-326 (which was confirmed later by Miller 1996) 
and pointed out that
the low optical surface brightness and blue color of this galaxy
is difficult to reconcile using current evolutionary models of late-type
galaxies.
However, it is shown here that the blue color of this galaxy is due to
a presence of young stellar populations. 

We have investigated roughly the properties of young stellar populations seen
in Fig. 2 using the theoretical isochrones.
In Fig. 8, we overlay, in the $V-(B-V)$ diagram of UKS 2323--326, three isochrones
for the metallicity of $Z=0.001$ ([Fe/H] = -- 1.2 dex) and ages of 
10, 30, and 100 Myrs given by the Padova Group (\cite{ber94}).
Fig. 8 shows (1) that the  brightest blue and red  stars in UKS 2323--326 match roughly
the blue and red loops of the isochrone with an age of 30 Myrs, and
(2) that the  brightest main-sequence at $V\approx 21.5$ 
(about 1 mag below the brightest blue supergiants) indicates an age
of 10 Myrs.
These results show that stars were formed as recently as 10 Myrs ago in 
UKS 2323--326.
On the other hand, a well-developed RGB shown in the $I-(V-I)$ diagram
 of Fig. 3
shows that the bulk of the stars in UKS 2323--326 were formed 
before a few Gyrs ago, and a small number of AGB stars above the TRGB
were probably formed a few Gyrs ago.

Fig. 9 displays the $V$ luminosity function of the main-sequence stars
with $(B-V)<0.4$ in UKS 2323--326. 
The luminosity function we derived is incomplete for the faint end, 
but is reasonably complete for $V<23$ mag. The bright part of the
luminosity function ($21.4<V<23$ mag, $-4.2 < M_V < -3.6$ mag) 
is approximately fit by a line 
with a logarithmic slope of $0.52\pm 0.12$, which is similar to those of
other irregular and spiral galaxies (\cite{fre86}, \cite{hoe86}).

The integrated HI flux of this galaxy was measured to be $15\pm3$ Jy \kms
 (\cite{lon78}, \cite{lon82}), and   Longmore \etal (1978) 
pointed out that the ratio of the HI mass to $B$ luminosity of this galaxy
they derived, $M_{HI} / L_B = 2.7~ M_\odot / L_\odot$, is  rather large for
normal irregular galaxies.
With better data for the distance and luminosity obtained in this study, 
we derive an HI mass, $M_{HI} = 1.5 \times 10^7$ $M_\odot$, 
a $B$ luminosity, $L_B = 1.67 \times 10^7 $ $L_\odot$, 
and a ratio of $M_{HI} / L_B = 0.9~ M_\odot / L_\odot$ which is much lower
than the value given by Longmore \etal (1978). 
Our value for the ratio of the HI mass to $B$ luminosity is typical for dwarf irregular
galaxies (\cite{mat98}).
 
\subsection{The Brightest Blue and Red Stars in UKS 2323--326}

Longmore \etal (1978) estimated roughly from the survey plate the magnitude
of the
brightest blue stars in UKS 2323--326 to be $B=19.3\pm 0.5$ mag, and used
this result to derive the distance to this galaxy, obtaining a value of
1.3 Mpc. With our photometry we can investigate in detail
the properties of the brightest stars in this galaxy.

In the $V-(B-V)$ diagram of the C-region of UKS 2323--326 (filled circles)
 shown in Fig. 2,
it is obvious which stars are the three brightest blue and red stars in UKS 2323--326. 
Three brightest blue stars (called BSG) are IDs 1055, 1002, and 1116, 
and three brightest red stars (called RSG) are IDs 956, 1058, and 821, 
as listed in Table 2. 
There is one very blue star (ID 1140) as bright as ID 1116, but the 
 color is too blue, $(B-V) = -0.49$ to be a normal star. Inspection of
the image of this object shows that the image is slightly asymmetric so
that the point-spread function fitting photometry of this object may not be 
as good as that of other normal stars. So we did not include it in the sample
of the three brightest blue stars. 
The mean magnitudes and colors of these three brightest blue and red stars 
in UKS 2323--326 are derived to be, respectively, 
$<V(3)>_{BSG} = 20.33\pm0.25$ mag and $<(B-V)(3)>_{BSG} = 0.14\pm0.07$,
and  
$<V(3)>_{RSG} = 20.74\pm0.18$ mag and $<(B-V)(3)>_{RSG} = 1.35\pm0.08$.
(If ID 1116 is replaced by ID 1140, the mean values will be slightly changed:
 $<V(3)>_{BSG} = 20.32\pm0.28$ mag and $<(B-V)(3)>_{BSG} = -0.04\pm0.32$.)
The corresponding absolute magnitudes and colors are
$<M_V(3)>_{BSG} = -6.31\pm0.25$ mag,  $<(B-V)(3)>_{BSG, 0} = 0.14\pm0.07$,
$<M_V(3)>_{RSG} = -5.91\pm0.18$ mag ,and $<(B-V)(3)>_{RSG, 0} = 1.34\pm0.08$, respectively.

The luminosity of the brightest stars in galaxies is known to be correlated
with the luminosity of the parent galaxies. 
Recently Lyo \& Lee (1997) presented, from the analysis of 17 galaxies 
(with $M_B <-14$ mag) to which Cepheid distances are available,
calibrations for the relation between the magnitudes of the brightest stars and 
the magnitudes of the parent galaxies:
$<M_V(3)>_{BSG} = 0.30 M_B ({\rm gal}) - 3.02$ with $\sigma = 0.55$ mag,
and $<M_V(3)>_{RSG} = 0.21 M_B ({\rm gal}) - 3.84$ with $\sigma = 0.47$ mag.
Using these relations, 
we derive $<M_V(3)>_{BSG} = -6.79$ mag and $<M_V(3)>_{RSG} = -6.48$ mag for the
absolute magnitude of UKS 2323--326 as derived in the previous section. 
Thus the magnitudes of the brightest stars in UKS 2323--326 are 
0.5--0.6 mag fainter than those
expected from the relation for the bright galaxies.

\section{SUMMARY AND CONCLUSION}

We have presented a study of the stellar populations in the dwarf irregular
galaxy UKS 2323--326
 based on deep $BVRI$ CCD photometry.
The primary results obtained in this study are summarized as follows and
the basic information of UKS 2323--326 is listed in Table 4.

(1) $BVRI$ color-magnitude diagrams of the stars in the $7'.35\times 7'.35$
 area of UKS 2323--326 have been presented.
 These color-magnitude diagrams  exhibit a blue plume, 
a well-defined RGB, and a small number of AGB stars with intermediate age.

(2) The tip of the RGB is found to be at $I=22.65\pm 0.10$  mag 
and $(V-I)=1.43\pm 0.05$ mag.
From this value we derive a distance modulus of UKS 2323--326 of $(m-M)_0=26.59\pm 0.12$ mag, 
and a distance of $2.08\pm 0.12$ Mpc. 
The corresponding distance of UKS 2323--326
from the center of the Local Group is derived to be 1.92 Mpc.
From this result and the systemic velocity of UKS 2323--326 ($v = 62 $ \kms ),
we conclude that UKS 2323--326 is outside the Local Group and is a member of the
Sculptor group.

(3) The mean color of the RGB at $M_I=-3.5$ mag is 
$(V-I)=1.29\pm 0.04$ mag.
 From this value we obtain a mean metallicity of the RGB:
[Fe/H] = $-1.98\pm0.17$ dex. The metallicity of the RGB in UKS2323--326
is close to the lowest end in the metallicity of dwarf irregular galaxies.

(4) The total magnitudes of UKS 2323--326 within $r_H$ are derived to be
  $M_B = -12.58$ mag, $M_V = -13.14$ mag, $M_R = -13.45$ mag, 
and $M_I = -13.79$ mag.
The central surface brightness is measured to be 
$\mu_B (0) = 23.4 $ mag arcsec$^{-2}$ and 
$\mu_V (0) =22.9  $ mag arcsec$^{-2}$.
Surface brightness profiles of the central part of UKS 2323--326 are 
approximately 
consistent with a King model with a core concentration parameter
$c = log (r_t / r_c ) \approx 1.0$, and those of the outer part
follow an exponential law. 

(5) The magnitudes of three brightest blue and red stars in UKS 2323--326
are derived: 
$<M_V(3)>_{BSG} = -6.31\pm0.25$ mag and $<M_V(3)>_{RSG} = -5.91\pm0.18$ mag,
 which are about half magnitude fainter than those 
expected from conventional correlations with galaxy luminosity.

\acknowledgments

The authors are grateful to Eunhyeuk Kim for providing the
surface photometry program, and to Sang Chul Kim for reading the manuscript
carefully.
This research is supported by
the Ministry of Education, Basic Science Research Institute grant 
No.BSRI-98-5411 (to M.G.L.) and
by Creative Research Initiatives Program of the Korean Ministry
of Science and Technology and also by Yonsei University Research Grant (to Y.I.B.). 

%

%
%

\clearpage


\begin{figure}[1] 
\plotone{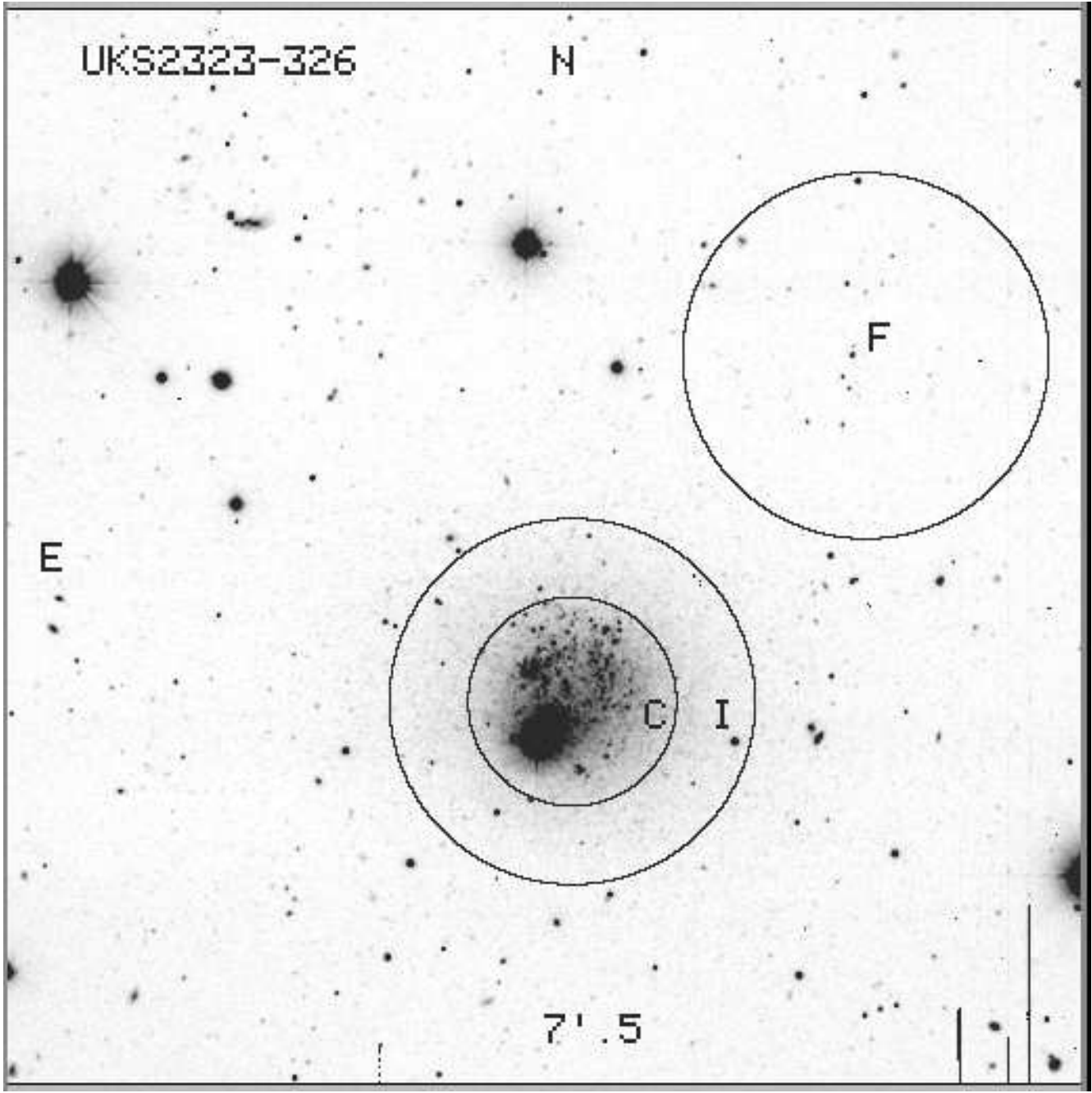}
\figcaption{
A greyscale map of the $V$-band CCD image of UKS 2323--326. 
North is at the top and east is to the left.
The size of the field is $7'.5 \times 7'.5$. 
Regions labelled as C, I, and F represent,
 respectively,
the central region ($r<44''$), the intermediate region ($44''<r<77''$) and the control field region ($r<77''$).
}
\end{figure}

\begin{figure}[2] 
\plotone{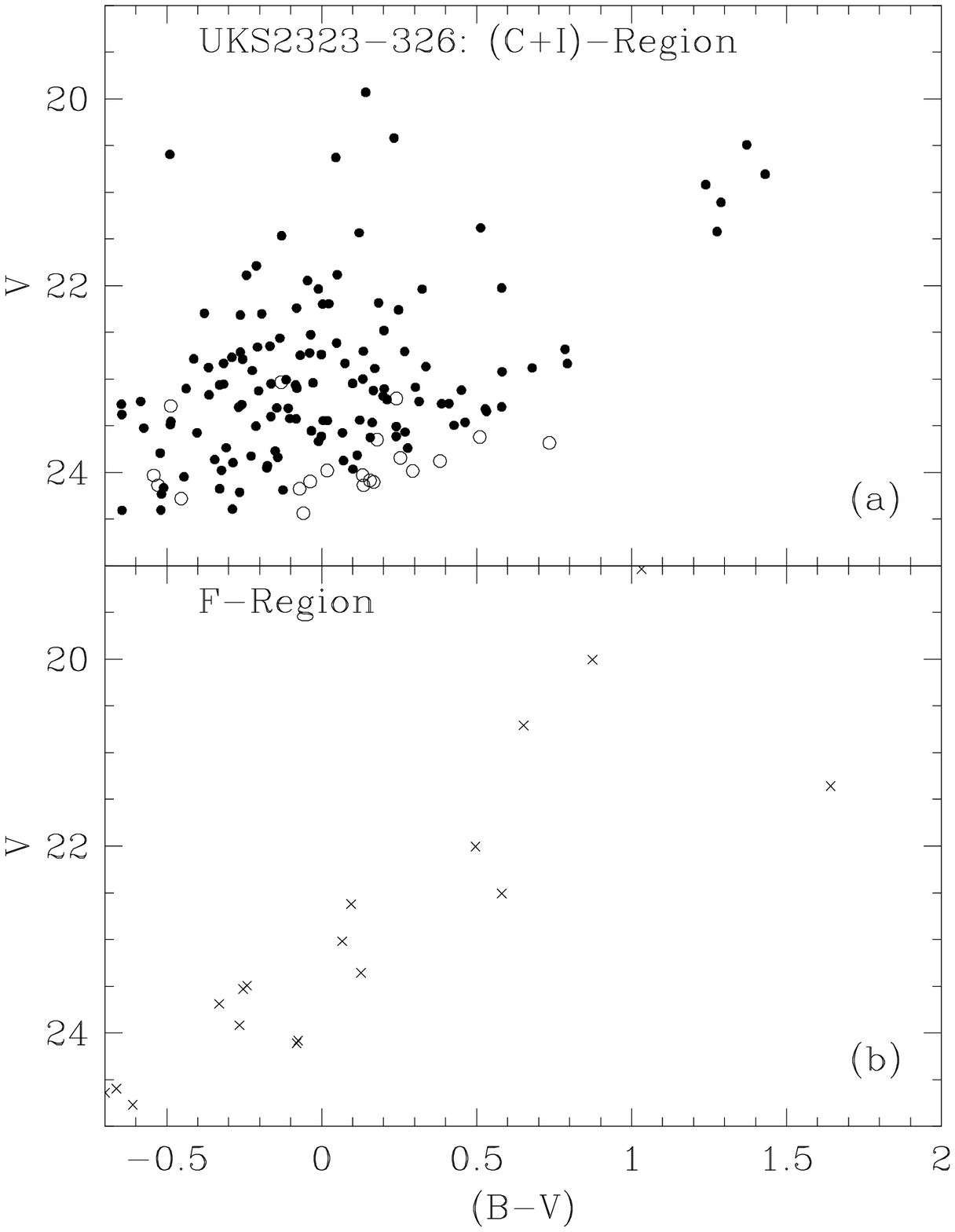}
\figcaption{(a) $V$--$(B-V)$ diagram of the measured stars in the C-region plus I-region of UKS 2323--326. 
Filled circles and open circles represent the stars in the C-region 
and I-region,  respectively.
(b) $V$--$(B-V)$ diagram of the measured stars in the F-region.}
\end{figure}

\begin{figure}[3] 
\plotone{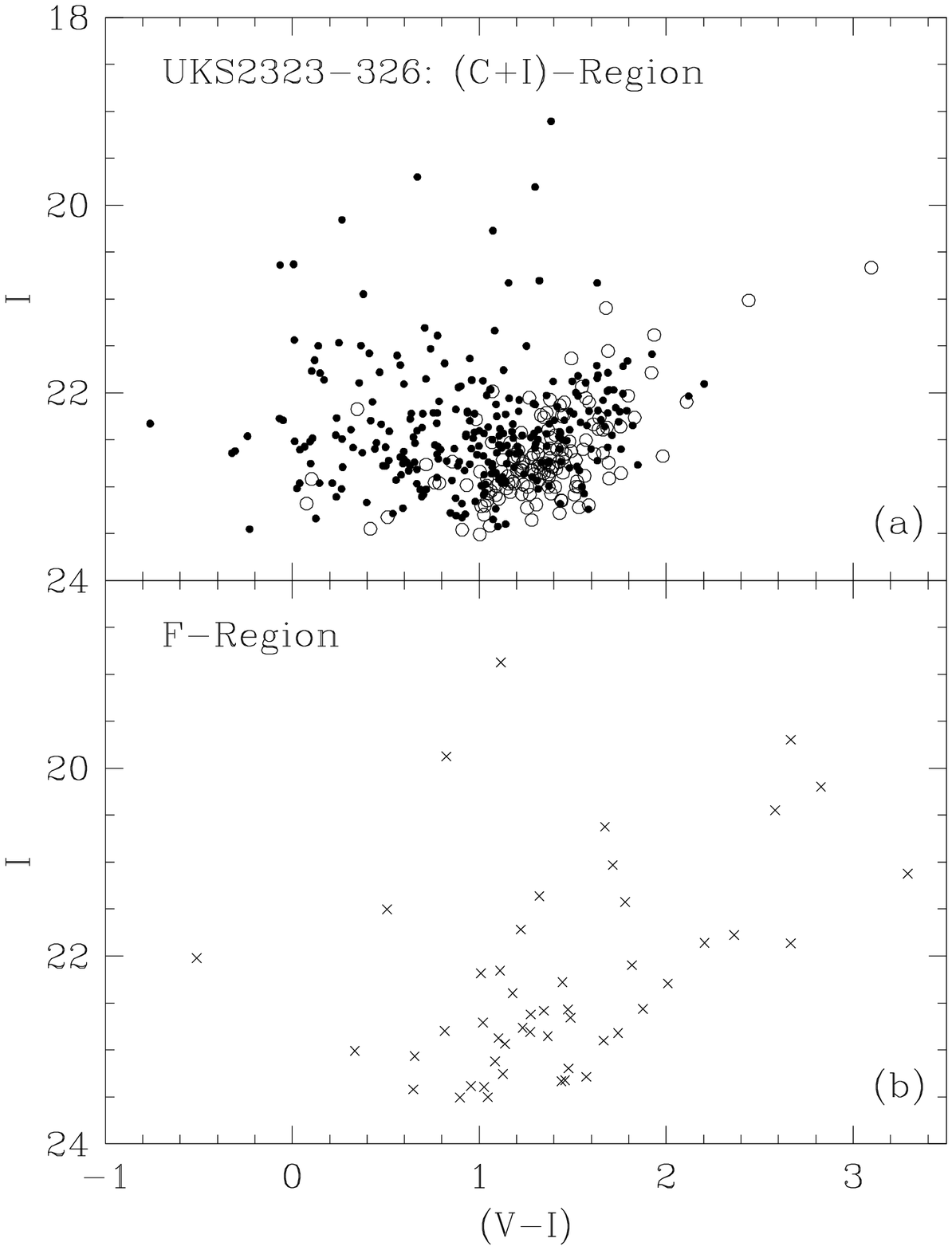}
\figcaption{
(a) $I$--$(V-I)$ diagram of the measured stars in the C-region plus I-region
of UKS 2323--326.
Filled circles and open circles represent the stars in the C-region 
and I-region, respectively.
(b) $I$--$(V-I)$ diagram of the measured stars in the F-region.
}
\end{figure}

\begin{figure}[4] 
\plotone{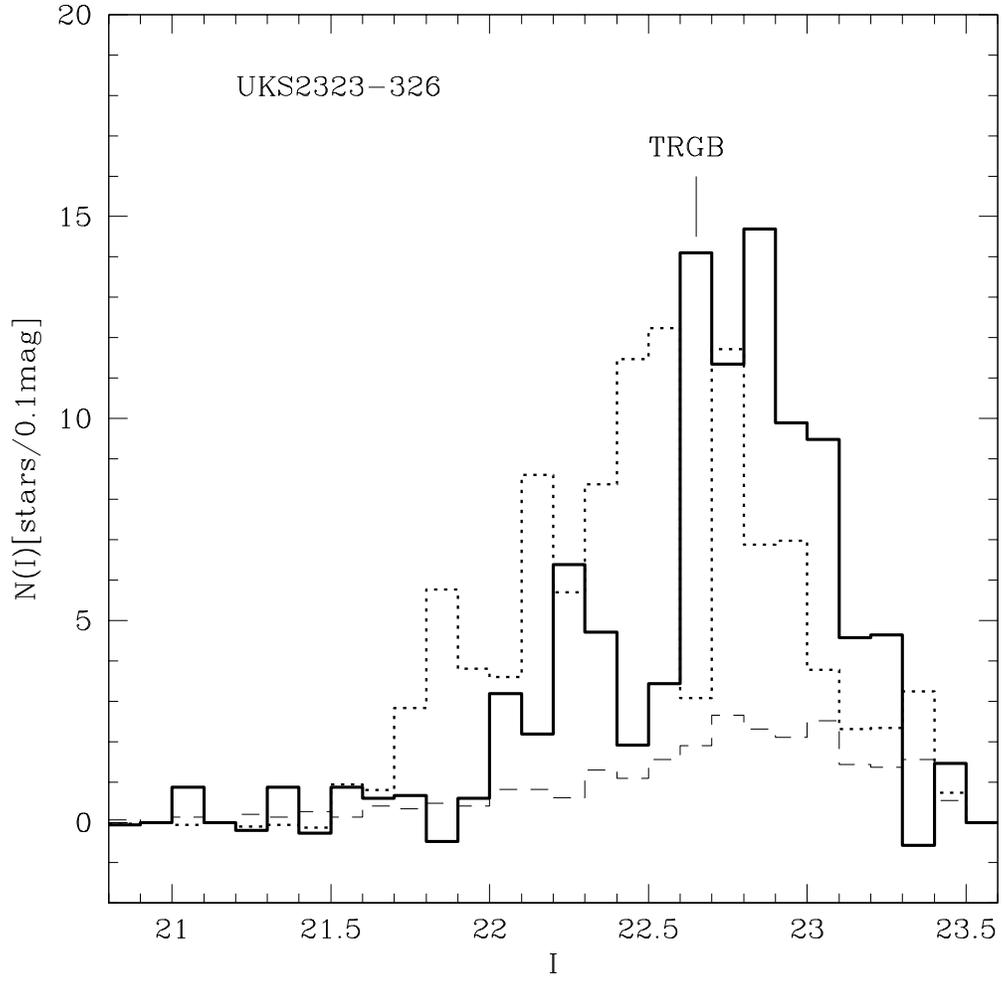}
\figcaption{
$I$-band luminosity function of the red giant branch stars in the C-region
and I-region. The tip of the red giant branch is labelled as TRGB.
The thick solid line, dotted line and dashed lines represent the luminosity functions of the I-region, C-region, and F-region, respectively.
The contribution due to field stars was subtracted from the luminosity
functions of the C-region and the I-region. }
\end{figure}

\begin{figure}[5] 
\plotone{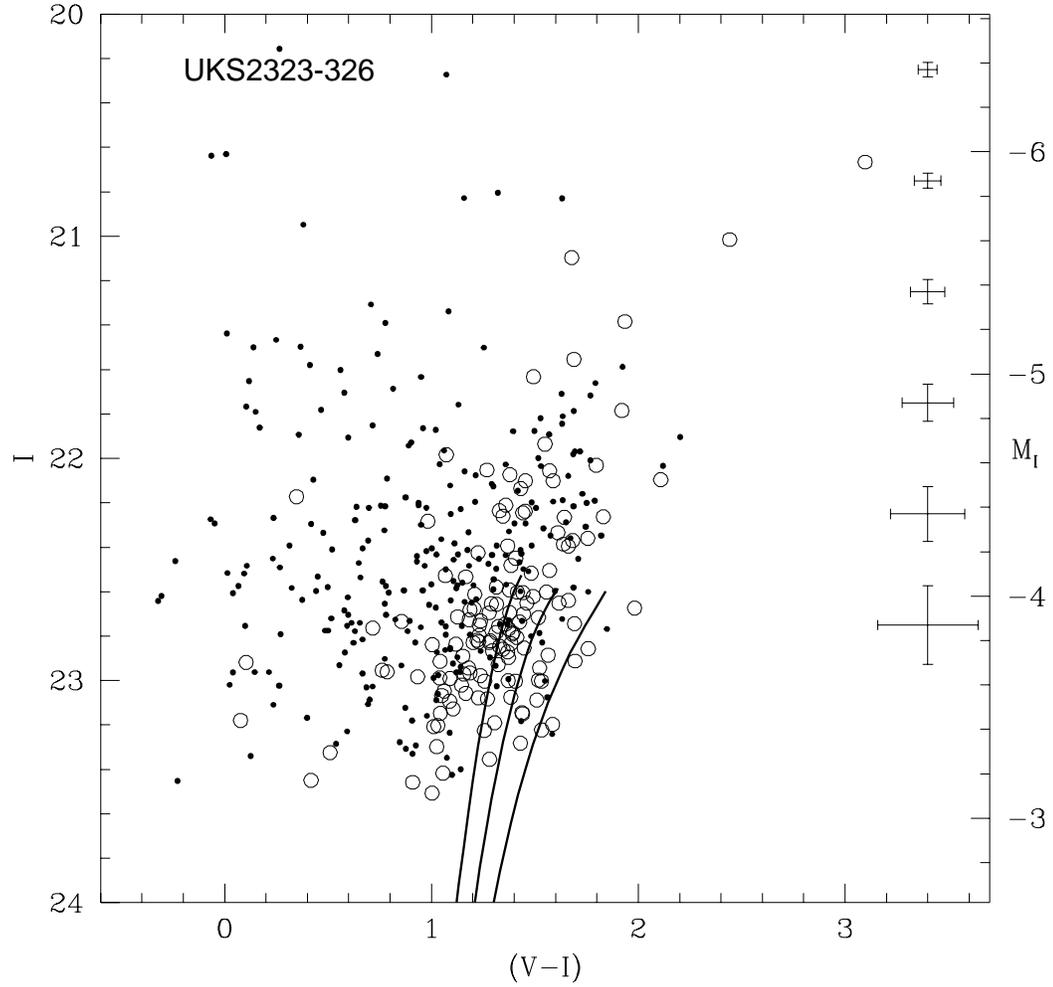}
\figcaption{
$I$--$(V-I)$ diagram of the measured stars in the I-region (open circles) and
C-region (filled circles) of UKS 2323--326 in comparison with
the red giant branches of Galactic globular clusters. 
The solid curved lines show,
from left to right, the loci of the giant branches of M15, M2, and NGC 1851, 
the metallicities of which are [Fe/H] = --2.17, --1.58 and --1.29 dex, respectively.
The mean errors for the magnitudes and colors are illustrated 
by the error bars at the right.
}
\end{figure}

\begin{figure}[6] 
\plotone{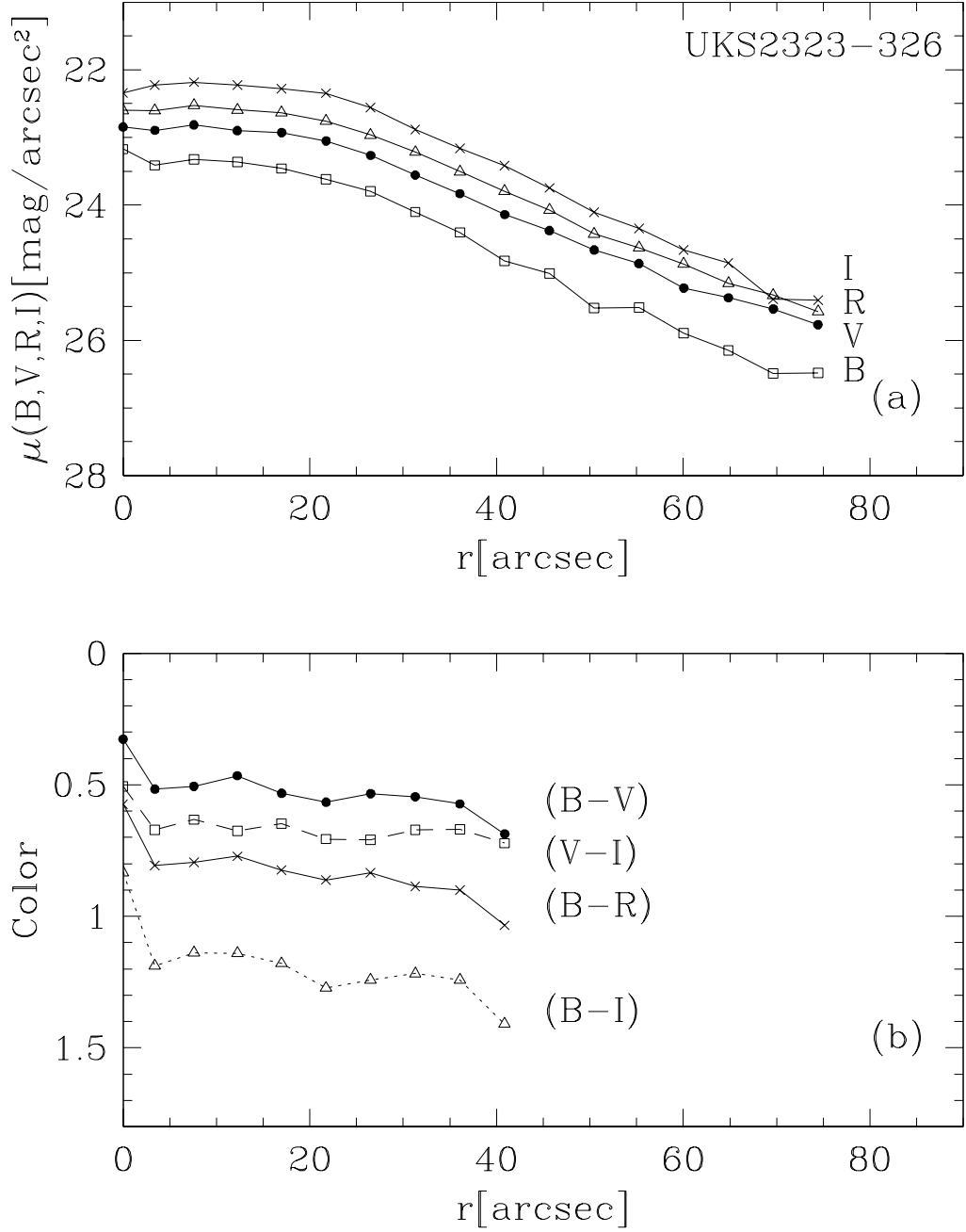}
\vskip-0.5in
\figcaption{ Surface photometry of UKS 2323--326.
(a) Surface brightness profiles vs  radius along the major axis.
$B,V, R$ and $I$ magnitudes are represented by the open squares, 
filled circles, open triangles, 
and crosses. 
(b) Differential colors vs  radius along the major axis.
}
\end{figure}

\begin{figure}[7] 
\plotone{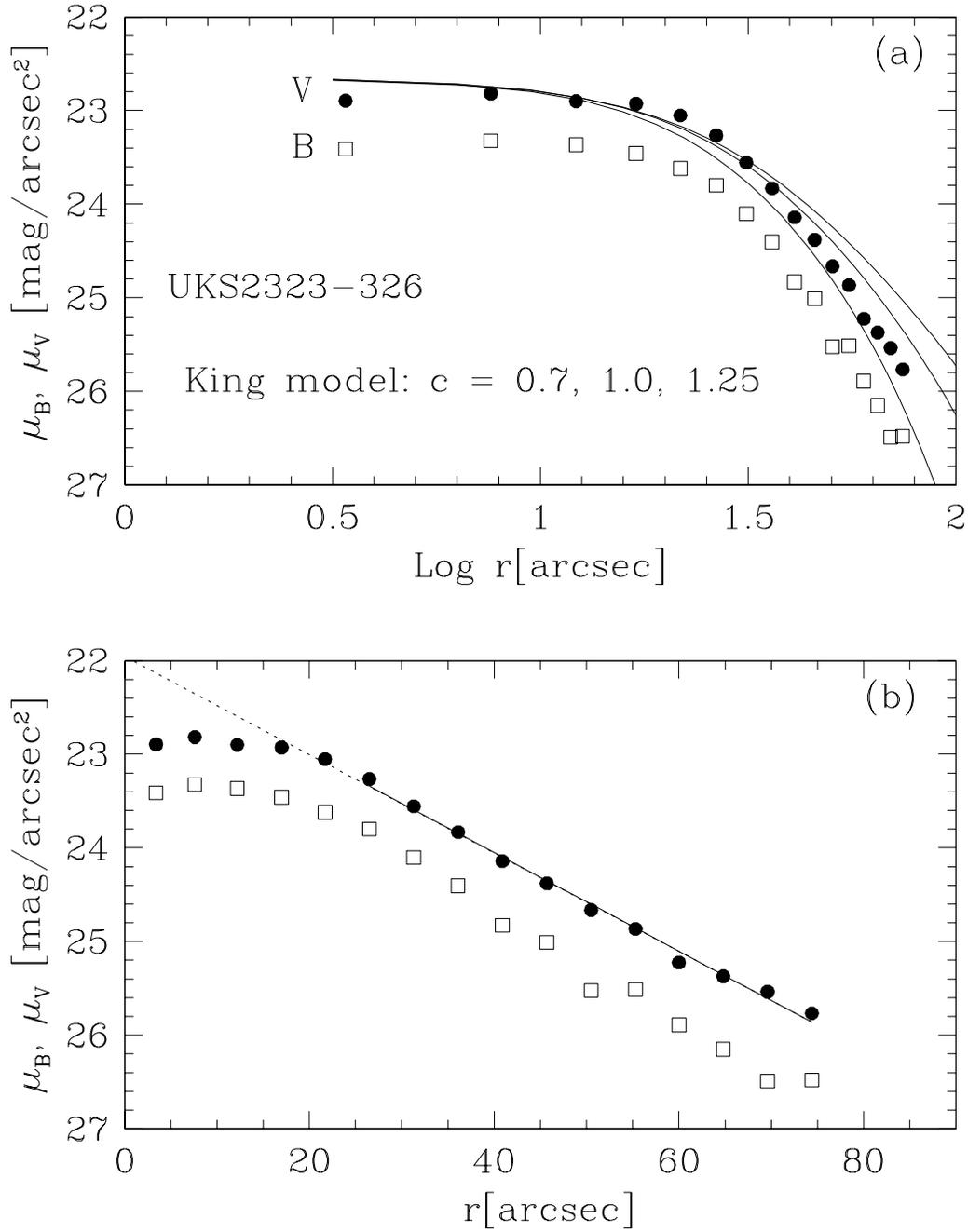}
\figcaption{Fits to the surface brightness profiles of UKS 2323--326.
(a) King model fitting. Filled circles and open squares represent,
respectively, $V$ and $B$ surface brightness profiles.
(b) Exponential law fitting.
}
\end{figure}

\begin{figure}[8] 
\plotone{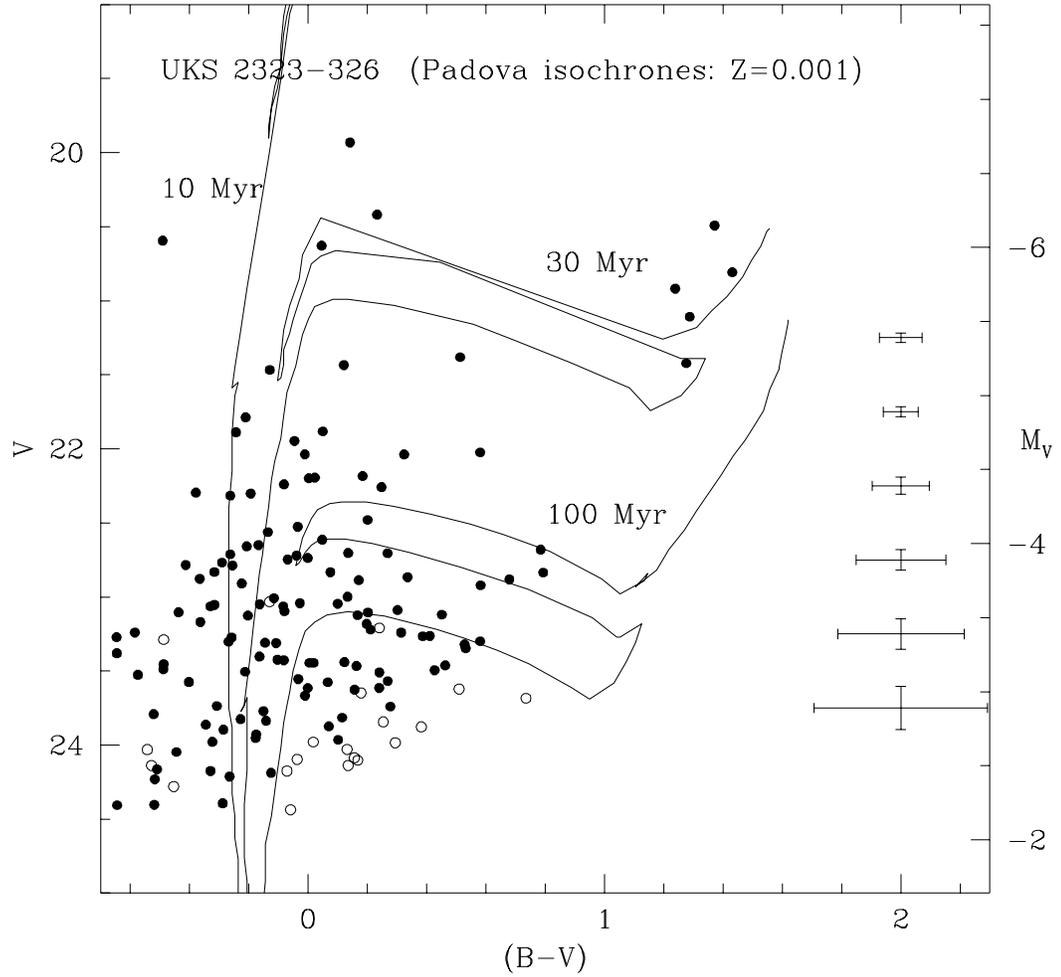}
\figcaption{
$V$--$(B-V)$ diagram of the measured stars in the C-region (filled circles) and
I-region (open circles) of UKS 2323--326 in comparison with theoretical isochrones.
The solid lines represent the Padova isochrones for the metallicity of $Z=0.001$
([Fe/H] = --1.2 dex) and ages of 10, 30, and 100 Myrs.
The mean errors for the magnitudes and colors are illustrated by the error bars
at the right.
}
\end{figure}

\begin{figure}[9] 
\plotone{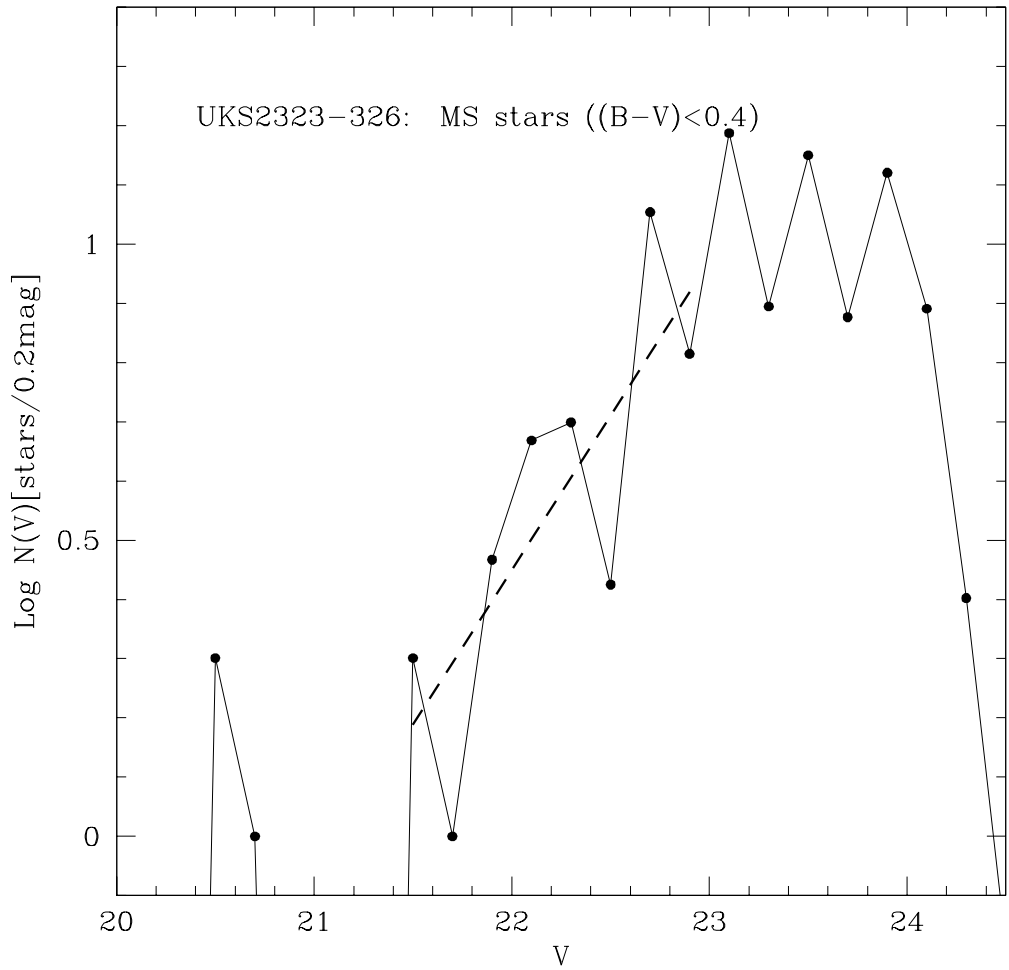}
\figcaption{ 
$V$-band luminosity function of the bright main-sequence stars with $(B-V)<0.4$
in UKS 2323--326. The dashed line represents a linear fit to the data for the
range of $21.4<V<23$ mag.}
\end{figure}

\clearpage


%








\begin{deluxetable}{ccccc}
\tablecaption{JOURNAL OF OBSERVATIONS FOR UKS 2323--326. \label{tbl-1}}
\tablehead{
\colhead{Filters} & \colhead{T$_{\rm exp}$}   & \colhead{Air mass}
& \colhead{FWHM} & \colhead{U.T.(Start)} } 
\startdata
$B$   &  300 s            & 1.68 & 1.1$''$ & 1994 Oct 7 09:41  \nl
$V$   &  $3 \times 700$ s & 1.66 & 1.0$''$ & 1994 Oct 7 09:00  \nl
$R$   &  $3 \times 500$ s & 1.63 & 1.0$''$ & 1994 Oct 7 08:27 \nl
$I$   &  $4 \times 400$ s & 1.66 & 0.9$''$ & 1994 Oct 7 07:48 \nl
\enddata
 
\end{deluxetable}





\begin{table*}
\begin{center}
\centerline{T{\small ABLE} 2.}
\centerline{PHOTOMETRY OF THE BRIGHT STARS WITH $V<22.5$ MAG IN  UKS2323--326.}
\bigskip
\begin{tabular}{ccccccc | ccccccc}
\tableline
\tableline
ID & X(px) & Y(px) & $V$ & ($B$--$V$) & ($V$--$R$) & ($V$--$I$) &
ID & X(px) & Y(px) & $V$ & ($B$--$V$) & ($V$--$R$) & ($V$--$I$) \nl
\tableline
    173&  810.9&  196.6&  17.83&   0.63&   0.41&   0.82&  1013&  532.0&  626.7&  21.43&   0.12&  --0.10&   0.01\nl 
  1659&  177.2&  805.2&  18.01&   0.56&   0.36&   0.69&  1445&  856.5&  713.0&  21.48&   0.79&   0.52&   1.02\nl 
  1858&  268.6&  921.5&  18.09&   0.65&   0.48&   0.89&   185&  805.8&  203.7&  21.60&   1.17&  --0.84&  --0.90\nl 
  1840&  661.3&  904.0&  18.50&   0.75&   0.56&   1.00&  1251&  445.4&  664.5&  21.79&  -0.21&  --0.10&   0.12\nl 
  1014&  520.5&  626.8&  18.93&   0.47&   0.76&   1.09&  1153&  924.3&  646.8&  21.86&   1.30&  --0.70&  --0.01\nl 
   132&  212.2&  163.2&  19.03&   1.03&   0.75&   1.40&   612&  121.4&  526.7&  21.88&   1.39&   0.91&   1.93\nl 
  1055&  532.2&  632.3&  19.93&   0.14&  --0.03&      &   997&  452.5&  624.0&  21.88&   0.05&   0.25&   0.37\nl 
   332&  217.6&  329.1&  20.00&   0.87&   0.62&   1.12&   999&  510.9&  624.3&  21.89&  --0.24&   0.06&   0.10\nl 
  1899&  205.9&  959.4&  20.35&   1.13&   0.83&   1.51&  1134&  490.3&  643.0&  21.95&  --0.05&  --0.14&   0.15\nl 
  1614&  270.6&  775.3&  20.38&   0.94&   0.72&   1.28&   225&  179.7&  234.9&  22.01&   0.50&   0.27&   0.51\nl 
    47&  583.7&   58.8&  20.42&   0.76&   0.41&   0.83&  1047&  459.0&  630.9&  22.02&   0.58&   0.70&   1.16\nl 
  1002&  522.6&  624.7&  20.42&   0.23&   0.04&   0.27&   980&  461.8&  621.8&  22.04&  -0.01&  --0.15&   0.17\nl 
   956&  508.3&  619.5&  20.49&   1.37&   0.81&   1.38&   885&  462.8&  606.8&  22.04&   0.32&   0.34&   0.71\nl 
  1140&  512.3&  644.1&  20.59&  --0.49&  --0.03&  --0.06&   930&  750.6&  615.1&  22.11&   0.72&  --0.20&  --0.22\nl 
  1896&  191.1&  953.4&  20.60&   0.52&   0.28&   0.67&   335&  667.4&  330.6&  22.15&   1.26&   1.95&   2.46\nl 
  1116&  508.8&  640.8&  20.63&   0.05&  --0.03&   0.01&   963&  452.1&  619.9&  22.18&   0.18&   0.33&   0.56\nl 
   251&  222.9&  261.1&  20.71&   0.65&   0.42&   0.82&  1483&  473.6&  725.3&  22.19&   0.02&  --0.01&  --0.07\nl 
  1058&  521.9&  632.7&  20.81&   1.43&   0.87&       &  1863&  555.2&  928.1&  22.20&   0.02&   0.25&   0.39\nl 
   821&  489.3&  593.8&  20.92&   1.24&   1.00&       &   968&  506.1&  620.1&  22.20&   0.00&  --0.30&  --0.24\nl 
    96&  813.8&  127.9&  21.03&   0.34&   0.18&   0.56&   900&  512.0&  609.4&  22.24&  --0.08&  --0.09&  --0.05\nl 
   234&  826.4&  247.7&  21.11&   1.45&   0.95&   2.15&  1691&  359.6&  818.7&  22.25&   0.52&   0.20&   0.84\nl 
   174&  740.0&  198.1&  21.18&   0.47&   0.35&   0.67&  1415&  449.9&  701.5&  22.26&   0.25&   0.10&   0.47\nl 
   975&  957.8&  620.9&  21.20&   0.30&   0.30&   0.61&   554&  925.4&  496.3&  22.38&   0.90&   0.25&   0.75\nl 
    21&  473.4&   27.0&  21.26&   1.56&   0.35&   0.16&   406&  714.3&  369.2&  22.40&   0.88&   0.14&   0.59\nl 
  1280&  166.7&  669.7&  21.30&   0.23&   0.17&   0.35&   388&  820.7&  359.4&  22.47&   0.03&   0.41&   0.38\nl 
   404&  932.9&  368.8&  21.34&   0.61&   0.29&   0.70&  1024&  486.6&  627.7&  22.48&   0.20&   0.62&   1.08\nl 
  1490&  477.1&  726.3&  21.38&   0.51&   0.75&   1.07&  1609&  241.3&  773.7&  22.49&   0.70&   1.49&   0.82\nl 
 \tableline

\end{tabular}
\end{center}
\end{table*}



\begin{deluxetable}{rcccc  rcccc}
\tablenum{3}
\tablecaption{$BVRI$ SURFACE PHOTOMETRY OF UKS 2323--326.
 \label{tbl-3}}
\tablewidth{0pt}
 \tablehead{ 
\colhead{$r_{\rm eff}$ [$''$]} & \colhead{$\mu_B$} & \colhead{$\mu_V$} &  \colhead{$\mu_R$} &\colhead{$\mu_I$} &
\colhead{$r_{\rm out}$ [$''$]} & \colhead{$B$} & \colhead{$V$} & \colhead{$R$} & \colhead{$I$}  }
\startdata 
  3.4   & 23.41   & 22.90   & 22.60   & 22.22  &  4.8   & 18.73   & 18.17   & 17.84   & 17.45 \nl
  7.6   & 23.32   & 22.82   & 22.53   & 22.18  &  9.6   & 17.05   & 16.58   & 16.28   & 15.93  \nl
 12.2   & 23.36   & 22.90   & 22.59   & 22.22  & 14.4   & 16.20   & 15.73   & 15.44   & 15.09 \nl
 17.0   & 23.46   & 22.93   & 22.64   & 22.28  & 19.2   & 15.62   & 15.14   & 14.85   & 14.47 \nl
 21.7   & 23.62   & 23.05   & 22.76   & 22.35  & 24.0   & 15.16   & 14.68   & 14.36   & 13.99  \nl
 26.5   & 23.80   & 23.27   & 22.96   & 22.56  & 28.8   & 14.83   & 14.36   & 14.04   & 13.67 \nl
 31.3   & 24.10   & 23.56   & 23.22   & 22.88  & 33.6   & 14.63   & 14.14   & 13.82   & 13.44 \nl
 36.1   & 24.40   & 23.83   & 23.50   & 23.16  & 38.4   & 14.48   & 13.97   & 13.65   & 13.26 \nl
 40.9   & 24.83   & 24.14   & 23.79   & 23.42  & 43.2   & 14.37   & 13.85   & 13.53   & 13.14 \nl
 45.7   & 25.01   & 24.38   & 24.07   & 23.75  & 48.0   & 14.28   & 13.76   & 13.43   & 13.06 \nl
 50.5   & 25.52   & 24.66   & 24.43   & 24.11  & 52.8   & 14.22   & 13.69   & 13.36   & 12.99 \nl
 55.3   & 25.51   & 24.87   & 24.63   & 24.35  & 57.6   & 14.17   & 13.63   & 13.30   & 12.94 \nl
 60.0   & 25.89   & 25.23   & 24.87   & 24.66  & 62.4   & 14.13   & 13.58   & 13.25   & 12.90 \nl
 64.8   & 26.15   & 25.37   & 25.16   & 24.86  & 67.2   & 14.10   & 13.53   & 13.21   & 12.86 \nl
 69.6   & 26.49   & 25.54   & 25.33   & 25.39  & 72.0   & 14.07   & 13.50   & 13.18   & 12.83 \nl
 74.4   & 26.48   & 25.77   & 25.57   & 25.41  & 76.8   & 14.04   & 13.46   & 13.14   & 12.80 \nl
 79.2   & 26.46   & 26.05   & 25.88   & 25.84  & 81.6   & 13.98   & 13.40   & 13.10   & 12.76 \nl
\enddata
\end{deluxetable}

\begin{deluxetable}{lcl}
\tablenum{4}
\tablecaption{BASIC INFORMATION OF UKS 2323--326.
 \label{tbl-4}}
\tablewidth{0pt}
 \tablehead{ \colhead{Parameter} & \colhead{Information} & \colhead{Reference} }
\startdata 
$\alpha_{1950}$, $\delta_{1950}$ & $23^h 23^m 47^s.6$ ,$-32^\circ 39' 57''$ &  1 \nl
$l, b$ & 11.86 deg, --70.86 deg  & 1 \nl
HI heliocentric radial velocity, $v_\odot$ & $62 \pm 5$ km s$^{-1}$ & 1 \nl
Foreground reddening, $E(B-V)$ & 0.014 mag & 2 \nl
Distance & $(m-M)_0 = 26.59\pm 0.12$, $d = 2.08\pm 0.12$ Mpc & this study \nl
Central surface brightness & $\mu_B(0) = 23.4$ , $\mu_V(0) = 22.9$ mag arcsec$^{-2}$ & this study \nl
Core radius & 32 arcsec = 320 pc & this study \nl
Standard radius, $r_{25}$ & 46 arcsec = 460 pc & this study \nl
Holmberg radius, $r_H$  & 70 arcsec = 700 pc & this study \nl
Apparent total magnitude ($<r_H$) & $B=14.07$ mag, $V=13.50$ mag & this study \nl
Absolute total magnitude & $M_B = -12.58 $ mag, $M_V = -13.14$ mag & this study \nl
RGB metallicity, [Fe/H] & $-1.98\pm 0.17$ dex & this study \nl
HI Flux & $15\pm3$ Jy km$^{-1}$ & 1 \nl
HI mass, $M_{\rm HI}$  & $1.5\pm 0.3 \times 10^7 M_\odot $  & this study \nl
\enddata
\tablerefs{ (1) Longmore et al. (1978); (2) Schlegel, Finkbeiner \& Davis (1998).}
\end{deluxetable}

%


\clearpage


\end{document}